\documentclass[12pt,english]{article}
\usepackage{amssymb}
\usepackage{babel}
\usepackage{latexsym}
\usepackage{graphics}
\usepackage{amsmath}
\usepackage{amsfonts}
\usepackage{hyperref}
\makeatletter
\@addtoreset{equation}{section}
\makeatother

\def\dalemb#1#2{{\vbox{\hrule height .#2pt
        \hbox{\vrule width.#2pt height#1pt \kern#1pt
                \vrule width.#2pt}
        \hrule height.#2pt}}}

\def\0{{\sst{(0)}}}
\def\1{{\sst{(1)}}}
\def\2{{\sst{(2)}}}
\def\3{{\sst{(3)}}}
\def\4{{\sst{(4)}}}
\def\5{{\sst{(5)}}}
\def\6{{\sst{(6)}}}
\def\7{{\sst{(7)}}}
\def\8{{\sst{(8)}}}
\def\n{{\sst{(n)}}}

\let\a=\alpha \let\b=\beta \let\g=\gamma \let\d=\delta \let\e=\epsilon
\let\z=\zeta    \let\k=\kappa
\let\l=\lambda \let\m=\mu \let\n=\nu  \let\r=\rho
\let\s=\sigma \let\t=\tau  \let\f=\phi  
\let\w=\omega    \let\L=\Lambda
   \let\F=\Phi 
    \let\G=\Gamma

 \def\bd{\begin{document}} \def\ed{\end{document}}
\def\ds{\documentstyle} \let\fr=\frac \let\bl=\bigl \let\br=\bigr
\let\Br=\Bigr \let\Bl=\Bigl
\let\bm=\bibitem
\let\na=\nabla
\let\pa=\partial \let\ov=\overline
\newcommand{\be}{\begin{equation}}
\newcommand{\ee}{\end{equation}}
\def\ba{\begin{array}}
\def\ea{\end{array}}
\def\ft#1#2{{\textstyle{{\scriptstyle #1}\over {\scriptstyle #2}}}}
\def\fft#1#2{{#1 \over #2}}
\def\del{\partial}
\def\sst#1{{\scriptscriptstyle #1}}
 \def\oneone{\rlap 1\mkern4mu{\rm l}}
\def\ie{{\it i.e.\ }}
\def\via{{\it via}}
\def\semi{{\ltimes}}
\def\str{{\rm str}}
\def\Dm{{{D_{\sst{max}}}}}
\def\vac{ \left | 0 \right \rangle }
\def\kvac{ \left | k \right \rangle }

\def\sp{\; \; \;}

\def\bol{ \left | B (p^+) \right \rangle}
\def\bo1{ \left | B^0 (p^+) \right \rangle}

\def\bolt{ \left | B (p^+) \right \rangle_{\t}}

\def\boxl{ \left | B (x^-) \right \rangle}

\def\<{ \langle }
\def\>{ \rangle }

\def\vf{\varphi}

\def\ls{{(l,0)}}
\def\lv{{(l,\pm1)}}
\def\lt{{(l,\pm2)}}

\def\lse#1{{(l_{#1},0)}}
\def\lve#1{{(l_{#1},\pm1)}}
\def\lte#1{{(l_{#1},\pm2)}}

\def\lsg#1{{5(l_{#1},0)}}
\def\lvg#1{{5(l_{#1},\pm1)}}
\def\ltg#1{{5(l_{#1},\pm2)}}

\def\lsi#1{{5{(#1,0)}}}
\def\lvi#1{{5{(#1,\pm1)}}}
\def\lti#1{{5{(#1,\pm2)}}}

\def\lsr#1{{1{(#1,0)}}}
\def\lvr#1{{1{(#1,\pm1)}}}
\def\ltr#1{{1{(#1,\pm2)}}}

\def\cn{{\cal N}}
\def\cao{{\cal O}}
\def\cD{{\cal D}}
\def\cE{{\cal E}}
\def\cF{{\cal F}}
\def\cG{{\cal G}}
\def\cH{{\cal H}}
\def\cK{{\cal K}}
\def\cO{{\cal O}}
\def\cP{{\cal P}}
\def\cQ{{\cal Q}}
\def\cR{{\cal R}}
\def\cS{{\cal S}}
\def\cT{{\cal T}}
\def\cU{{\cal U}}
\def\cV{{\cal V}}
\def\cW{{\cal W}}

\newcommand{\nono}{\nonumber}
\newcommand{\dtilde}[1]{\tilde{\tilde{#1}}}
\newcommand{\hatb}[1]{\hat{\ov{#1}}}
\newcommand{\hatt}[1]{\hat{\tilde{#1}}}
\newcommand{\emnr}{{e_\m}^{\n\r}}

\newcommand{\hsp}{\hspace{0.5cm}}

\newcommand{\ho}[1]{$\, ^{#1}$}
\newcommand{\hoch}[1]{$\, ^{#1}$}
\newcommand{\bea}{\begin{eqnarray}}
\newcommand{\eea}{\end{eqnarray}}
\newcommand{\ra}{\rightarrow}
\newcommand{\lra}{\longrightarrow}
\newcommand{\Lra}{\Leftrightarrow}
\newcommand{\ap}{\alpha^\prime}
\newcommand{\bp}{\tilde \beta^\prime}
\newcommand{\tr}{{\rm tr} }
\newcommand{\Tr}{{\rm Tr} }
\newcommand{\NP}{Nucl. Phys. }

\newcommand{\ams}{{\it Institute for Theoretical Physics,
University of Amsterdam, \\
Valckenierstraat 65, 1018XE Amsterdam, The Netherlands} \\
{\tt joosth, skenderi@science.uva.nl}}
\newcommand{\auth}{Joost Hoogeveen and Kostas Skenderis}

\def\red{\color{red}}

\vspace{10pt}

\begin{document}

\begin{flushright}
\hfill{ITFA-2007-49}
\end{flushright}

\vspace{25pt}

\begin{center}

{\Large \bf BRST quantization of the pure spinor superstring}

\vspace{20pt}

\auth

\vspace{15pt}

\vspace{8pt}

{\ams}

\vspace{20pt}

\underline{ABSTRACT}
\end{center}

We present a derivation of the scattering amplitude
prescription for the pure spinor superstring from first principles,
both in the minimal and non-minimal formulations,
and show that they are equivalent. This is
achieved by first coupling the worldsheet action to topological gravity
and then proceeding to BRST quantize this system. Our analysis
includes the introduction of constant ghosts and
associated auxiliary fields needed to gauge fix
symmetries associated with zero modes. All fields
introduced in the process of quantization can be
integrated out explicitly, resulting in the prescriptions
for computing scattering amplitudes that
have appeared previously in the literature.
The zero mode insertions in the path integral
follow from the integration over the constant auxiliary fields.

\pagebreak

\tableofcontents
\addtocontents{toc}{\protect\setcounter{tocdepth}{2}}

\section{Introduction}

The covariant super-Poincar\'{e} quantization of the ten dimensional
superstring is an important problem that has attracted lots of attention over
the years. There are strong motivations for developing such a formalism.
To start with one would like to be able to compute
multi-loop amplitudes in a manifestly supersymmetric fashion
and analyze the associated issue of the perturbative
finiteness of the superstring perturbation theory. Furthermore, holographic
dualities and the study of flux vacua in string theory make
urgent the need for a formalism that can handle RR backgrounds.

A new formalism that achieves such a covariant quantization,
the pure spinor formalism, was proposed by Berkovits in
\cite{Berkovits:2000fe}, see \cite{Berkovits:2002zk} for a review.
The worldsheet fields, in the minimal version of this formalism,
are the spacetime coordinates $X^m$ and the
spacetime fermions $\theta^\a$, as in the Green-Schwarz formalism,
conjugate fermionic momenta $p_\a$ (first introduced
by Siegel  in \cite{Siegel:1985xj}) and
new bosonic twistor-like variables $\l^\a$  that take values
in the space of pure spinors, namely they satisfy
$\l^\a \g^m_{\a \b} \l^\b=0$,
and their conjugate momenta $w_\a$. The non-minimal version contains
an additional bosonic pure spinor of opposite chirality, $\bar{\l}_\a$,
a constrained fermionic spinor $r_\a$ and their conjugate
momenta, $\bar{w}^\a, s^\a$. One can construct from these fields a
fermionic nilpotent operator $Q_S$ that is postulated to be the BRST operator
of the theory.
In a flat background the worldsheet theory is
free (modulo the non-linear pure spinor constraint) and
a prescription for the computation of scattering amplitudes has been
developed in a number of papers
\cite{Berkovits:2000fe,Berkovits:2004px,Berkovits:2005bt,Berkovits:2006vi}
with tests and explicit computations presented in
\cite{Berkovits:2000ph,Berkovits:2005df,Berkovits:2005ng,Policastro:2006vt,Berkovits:2006bk}.

There are several unconventional aspects of this formalism. Usually the
BRST symmetry arises after gauge fixing a local symmetry, which in the
case of strings includes worldsheet diffeomorphisms. In the pure spinor
formalism however one instead is given directly the ``gauged-fixed''
model in the
conformal gauge and a BRST operator $Q_S$.
Similarly, the prescription for the scattering
amplitudes was postulated rather than derived from first principles.
In particular, the absence of an antighost field $b$ led to a
(complicated) construction of a composite field, with properties
similar to that of the antighost, which was used in the proposal
for the measure of the multi-loop amplitudes. Although there is very little
doubt that the current form of the computation rules is correct,
it would clearly be desirable to have a first principles
derivation. Such a derivation, apart from providing a
better justification of the current computational rules, could
also help in the search of a simplified version.

In this paper we provide such a derivation. There have been many works
in the past involving  modifications and/or extensions
of the pure spinor formalism with the same aim, see for example
\cite{Grassi:2001ug,Grassi:2003cm,Grassi:2003kq,Oda:2001zm,Matone:2002ft,Aisaka:2005vn,Berkovits:2004tw,Gaona:2005yw}.
Our approach is different and is guided by topological string constructions.
Instead of searching for a model with a local symmetry which after
gauge fixing would lead to the pure spinor formalism with
$Q_S$ and the
pure spinors emerging as a BRST operator and ghost fields,
we shall consider the pure spinors $\l$ as ``matter'' fields as well
and the worldsheet theory as a sigma model with a
nilpotent symmetry $Q_S$ and
target space ten dimensional superspace
times the pure spinor space. To construct a
string theory we couple this theory to two-dimensional gravity
in a way that preserves the fermionic symmetry $Q_S$ and then BRST quantize
the resulting theory in a conventional fashion. Following
\cite{Craps:2005wk}, gauge invariances due to zero modes
are also included in the BRST analysis by introducing
constant ghosts. This leads automatically to a scattering
amplitude prescription that is BRST invariant and upon
integrating out the constant ghosts and associated auxiliary fields
one arrives at various insertions in the path integral measure.

This paper is organized as follows. In the next section we
review the pure spinor formalism. In section \ref{cp2d} we
couple the theory to $2d$ topological gravity and in
section \ref{VO} we introduce vertex operators. The BRST quantization of this
system is presented in section \ref{BRST}. Section \ref{ps_measure}
contains the analysis of the invariances due to the
pure spinor zero modes. In this section we show that
depending on how one treats the auxiliary fields
one arrives either at the minimal or the non-minimal
scattering amplitude prescription. We conclude in section \ref{concl}
with a brief summary of our results.
Finally there are two appendices: in the first we discuss
$U(5)$ variables and the $Y$ formalism, while the second
contains details of computations relevant for section \ref{BRST}.

\section{Review of the pure spinor formalism} \label{review}

We review in this section the pure spinor formalism.
In the minimal version, the worldsheet action for the left-movers
in conformal gauge and flat target space is given
by
\be \label{fl_action}
S_\s= \int d^2z \left(\frac{1}{2}\partial x^m \bar{\partial}x_m
+p_{\alpha} \bar{\partial}\theta^{\alpha}
- w_{\alpha}\bar{\partial}\lambda^{\alpha}\right)
\ee
with $m=0,\ldots, 9$ and $\a=1,\ldots, 16$.
For Type II strings the right-movers are similar to the left-movers
while for the heterotic string the right-moving variables are those
of the heterotic RNS formalism.
The field $\l^\a$ is a bosonic pure spinor satisfying,
\be \label{ps}
\l^\a \g^m_{\a \b} \l^\b =0,
\ee
where  $\g^m_{\a \b}$ are the symmetric
$16 \times 16$ $d=10$ Pauli matrices.
The fields $\theta^\a, \l^\a$,
have conformal dimension 0 and the corresponding conjugate momenta
$p_\a, w^\a$ have conformal dimension 1.

Since the action is quadratic in fields quantization is straightforward
except for the fact that $\l^\a$ satisfies the quadratic constraint
(\ref{ps}), so its quantization requires some explanation.
More precisely, the pure spinor part of the action is
a curved $\beta \gamma$ system describing maps from the
worldsheet to the space of pure spinors, with $\l^\a$ being
holomorphic coordinates in this space.
This system can be analyzed by covering the space of pure
spinors with coordinate patches and then gluing appropriately
on the overlaps \cite{Witten:2005px,Nekrasov:2005wg}.
In particular, one may cover the pure spinor space
with 16 coordinate patches chosen such that
on each of them one of the pure spinor components $\l^\a$
is non-vanishing. On such a patch one may explicitly
solve the pure spinor constraint to express $\l^\a$
in terms of 11 independent (complex) variables.
For example, using the decomposition
$\underline{16} \to \underline{1}+\underline{10} + \underline{5}^*$
of the spinor of (the Wick rotated Lorentz group)
$SO(10)$ under $SU(5)$ one may
solve the pure spinor constraint
 by suitably expressing  the $\underline{5}^*$ in terms of
the $\underline{1}$ and $\underline{10}$.
Furthermore, the action (\ref{fl_action}) has a gauge invariance
\be \label{ginv}
\d w_\a = \L_m (\g^m \l)_\a \; ,
\ee
where $\L^m$ is a gauge parameter, which on each patch can be
used to eliminate 5 components of $w_\a$, so we are left with
11 conjugate momenta for the 11 independent pure spinor components.
In appendix \ref{u5} we show how to implement these steps
in the path integral starting from a Lorentz invariant action
involving unconstrained spinors $\l^\a$ and a Lagrange multiplier
$l_m$ that imposes the pure spinor condition. Integrating out the
Lagrange multiplier and the ghost fields resulting from gauge fixing
the invariance (\ref{ginv}) one obtains (after a non-trivial cancellation)
a free action for the 11 independent pure spinor variables and their
conjugate momenta. Since the starting point is Lorentz invariant
all computations done with the $U(5)$ variables will preserve Lorentz
invariance.

The model is invariant under a fermionic nilpotent symmetry
(which for the left-movers is)
generated by
\be \label{qs}
Q_S = \oint dz \l^\a(z) d_\a(z),
\ee
where
\be
d_\a = p_\a - \frac{1}{2} \g_{\a\b}^m  \theta^\b \pa x_m - \frac{1}{8}
\g_{\a \b}^m \g_{m \ \g \d} \theta^\b \theta^\g \pa \theta^\d,
\ee
which is considered to play the role of the BRST operator.
The transformations it generates are given by
\be \label{S-tr}
\delta_S x^m= \lambda \gamma^m \theta, \quad
\delta_S \theta^{\alpha}= \lambda^{\alpha}, \quad \delta_S \l^\a =0, \quad
\delta_S d_{\alpha}=-\Pi^m(\gamma_m \lambda)_{\alpha}, \quad
\delta_S w_{\alpha}= d_{\alpha},
\ee
where $\Pi^m = \pa x^m + \frac{1}{2} \theta \g^m \pa \theta$ is the
supersymmetric momentum and again we restrict to the left-movers
(so in particular,
the full transformation for $x^m$ contains a similar additive term
with right-moving fields).
The cohomology of this operator (at ghost number one) indeed correctly
reproduces the superstring spectrum \cite{Berkovits:2000nn}.

The non-minimal version of the formalism \cite{Berkovits:2005bt}
amounts to introducing a set of non-minimal variables,
the complex conjugate $\bar{\l}_\a$ of $\l^\a$, a fermionic
constrained spinor $r_\b$ satisfying
\be
\bar{\l}_\a \g_m^{\a \b} \bar{\l}_\b =0, \qquad
\bar{\l}_\a \g_m^{\a \b} r_\b =0,
\ee
and their conjugate momenta, $\bar{w}^\a$ and $s^a$.
The action (\ref{fl_action}) is modified by the
addition of the term $S_{nm}$
\be \label{nn_ac}
S_\s \to S_\s+ S_{nm}, \qquad S_{nm} = \int d^2 z \left(
-\bar{w}^{\alpha}\bar{\partial}\bar{\lambda}_{\alpha}
+s^{\alpha}\bar{\partial}r_{\alpha}\right),
\ee
and the generator $Q_S$ by
\be
Q_S \to Q_S + \oint dz \bar{w}^\a r_\a
\ee
This acts on the non-minimal variables as follows
\be \label{qs_nm}
\delta_S \bar{\lambda}_{\alpha}= r_{\alpha}, \qquad \delta_S r_{\alpha} =0,
\qquad \delta_S s^{\alpha}= \bar{w}^{\alpha}, \qquad
\delta_S \bar{w}^{\alpha} = 0.
\ee
These transformation rules imply that the cohomology is independent of
the non-minimal variables.

\section{Coupling to 2d gravity} \label{cp2d}

To construct a string theory we will couple the theory discussed in the
previous section to two-dimensional gravity in a way that preserves
the $Q_S$ symmetry and then quantize this system. Since this model
has zero central charge, one should
couple it to topological gravity. Our approach is thus similar
to the construction of topological string theories, see \cite{Dijkgraaf:1990qw}
for a review. In that
context one starts from a supersymmetric sigma model
which upon topological twisting yields a topological sigma model.
In this procedure one of the supersymmetry charges is identified
with the BRST operator of the sigma model. The corresponding operator
in our case is the nilpotent operator $Q_S$. Note that
the pure spinor sigma model has been obtained
by twisting an $N=2$ model in \cite{Matone:2002ft}.

The first step in this procedure is thus to relax the conformal
gauge in the action (\ref{fl_action}) (or (\ref{nn_ac}) for the non-minimal
version). The part that involves the $x^m$ is standard\footnote{We work
with an Euclidean worksheet and use standard conventions, i.e.
$z=\s^1 + i \s^2$, the flat metric is $g_{z \bar{z}}=1/2$ etc.},
\be
S_X = \int d^2 \s (\frac{1}{4} \sqrt{g} g^{ab} \pa_a x^m \pa_b x_m)
\ee
The rest of the action (\ref{fl_action}) (or (\ref{nn_ac}) for the non-minimal
version) is a sum of first order actions involving a field of dimension
one and a field of dimension zero (with an overall sign that depends
on whether the fields are bosonic or fermionic). The covariantization
of all these terms is the same, so it suffices to discuss one of them,
say
\be \label{pth}
S_{(p,\theta)} = \int d^2z p_\a \bar{\partial} \theta^\a\ .
\ee
The fields of dimension one are vectors on the worldsheet, so $p_\a$
is more accurately labeled as $p_{a\a}$. However, only the $z$-component
participates in (\ref{pth}). Similarly, only the $\bar{z}$ component
of the right-moving momentum\footnote{Note that throughout this
article we use the notation that right-moving fields have a tilde
(rather than the more conventional bar).}
 $\tilde{p}_{a \a}$ participates in the
action. To account  for this, we introduce the projection
operators
\be
P^{(\pm)b}_{\ \ a} = \frac{1}{2} (\d_a{}^b \mp i J_a{}^b)\, ,
\ee
where $J_a{}^b$ is the complex structure of the worldsheet, i.e.
it satisfies
\be \label{cpx}
J_a{}^b J_b{}^c = - \d_a^c, \qquad \nabla_c J_a{}^b =0.
\ee
In terms of the worldsheet volume form and the worldsheet metric,
it is given by
$J_a{}^b = - \e_{a c} g^{cb}$, with $\e_{a b} = \sqrt{g} \hat{\e}_{ab}$
and $\hat{\e}_{01}=1$, and holomorphic and anti-holomorphic functions on
the worldsheet are defined by  $J_a{}^b \pa_b f = i \pa_a f$ and
$J_a{}^b \pa_b \tilde{f} = -i \pa_a \tilde{f}$, respectively.
Using (\ref{cpx}) one shows that
\be
P^{(\pm)b}_{\ \ a} P^{(\pm)c}_{\ \ b} = P^{(\pm)c}_{\ \ a}
\qquad
P^{(\pm)b}_{\ \ a} P^{(\mp)c}_{\ \ b}= 0.
\ee
Notice also that
\be
g^{ab}  P^{(\pm)c}_{\ \ b} = g^{cb}P^{(\mp)a}_{\ \ b}\, .
\ee
One can obtain vectors with only
$z$-component by multiplying by $P^{(+)b}_{\ \ a}$
and vectors with only $\bar{z}$-component
by multiplying by $P^{(-)b}_{\ \ a}$:
\be
\hat{p}_a = P^{(+)b}_{\ \ a} p_b, \qquad
\hat{\tilde{p}}{}_a = P^{(-)b}_{\ \ a} \tilde{p}_b\, .
\ee
In other words, in complex coordinates
the only non-zero component of $P^{(+)b}_{\ \ a}$  is
$P^{(+)z}_{\ \ z}=1$
and  the only non-zero component of $P^{(-)b}_{\ \ a}$
is  $P^{(-)\bar{z}}_{\ \ \bar{z}}=1$.
More generally, these projection operators can be used
to covariantize any tensor given in conformal gauge.
The action (\ref{pth}) can then be covariantized as
\be
S_{(p,\theta)} = \int d^2 \s \sqrt{g} g^{ab}
\hat{p}_{a \a} \pa_b \theta^\a\, .
\ee

In summary the action of the minimal model coupled to gravity
is given by
\be
S_\s = \int d^2 \s \sqrt{g} g^{ab} \left(\frac{1}{4}
\pa_a x^m \pa_b x_m + \hat{p}_{a \a} \pa_b \theta^\a
- \hat{w}_{a \a} \pa_b \l^\a \right)
\ee
with an obvious addition for the case of the non-minimal model.
The stress energy tensor for the model can be obtained by
varying w.r.t. the worldsheet metric,
\bea \label{e/m}
T_{ab} &=& \frac{2}{\sqrt{g}} \frac{\d S_\s}{\d g^{ab}}
=\frac{1}{2}(\partial_a x_m \partial_b x^m-\frac{1}{2}g_{ab}g^{cd}
\partial_c x_m \partial_d x^m) \\
&+&(p_{(a |\alpha|}\partial_{b)} \theta^{\alpha}
-\frac{1}{2}g_{ab}g^{cd}p_{c \alpha}\partial_d \theta^{\alpha})
+ T^{(\lambda w)}_{ab}
\nonumber
\eea
The contribution of the pure spinor part (and the non-minimal
variables) is same as the one for the
$(p,\theta)$ part with $p \to w$ and $\theta \to \l$ and an
overall minus sign (with similar replacements for the
non-minimal fields). This stress energy tensor is (manifestly) traceless
and covariantly conserved, reflecting the fact that the action is
invariant under diffeomorphisms and Weyl transformations,
\bea \label{difW}
\delta g_{ab}&=&\mathcal{L}_{\epsilon(\sigma)}g_{ab}+2\phi(\sigma) g_{ab} \\
\delta \F&=&-\epsilon^a \partial_a \F \nonumber \\
\delta P_{a}&=&-\epsilon^a \partial_a P+\partial_a \epsilon^{b} P_{b} \nonumber
\eea
where $\e^a(\s),\f(\s)$ are diffeomorphism and Weyl gauge parameters,
$\mathcal{L}_{\epsilon}$ is the Lie derivative,
$\F=\{x^m, \theta^\a, \l^\a, \ldots\} $ denotes collectively all
worldsheet scalars and $P_a=\{p_{a\a}, w_{a\a}, \ldots \}$ denotes
collectively all worldsheet vectors.

The stress energy tensor (\ref{e/m}) can be rewritten as
\be \label{e/m1}
T_{ab} = P^{(+) c}_{\ \ a} P^{(+) d}_{\ \ b} T^B_{cd}
+ P^{(-)c}_{\ \ a} p_{c \a} \left(P^{(-)d}_{\ \ b} \pa_d \theta^\a\right)
+ \cdots
\ee
where the dots indicate the contribution from the pure spinor and
non-minimal variables, which will be suppressed from now on
since they are similar to the $(p,\theta)$ contribution.
We also suppress the anti-holomorphic contribution of $x^m$.
The first term in (\ref{e/m1}) is the covariantization
of the stress energy tensor appearing in Berkovits' work,
\be \label{TB}
T^B_{ab}=\frac{1}{2}\partial_a x_m \partial_b x^m
+p_{a \alpha} \partial_b \theta^{\alpha} + \cdots
\ee
while the second term is proportional to the $\theta^\a$ field equation.
This additional term can be removed by modifying the transformation
rule of $p_{a \a}$ in (\ref{difW}).

\subsection{Topological gravity and $Q_S$ invariance}

If we were to quantize the model just described we would find that
it is anomalous, since the diffeomorphism ghosts would
contribute $c=-26$ and the original sigma model had $c=0$.
This problem is avoided by extending the $Q_S$ symmetry to
act on the worldsheet metric, so that the $2d$ gravity is topological.
With this aim, we introduce the following transformation
rule,
\be \label{Sgab}
\d_S g_{ab} = P^{(-)c}_{\ \ a} P^{(-)d}_{\ \ b} \psi_{cd} \equiv
\hat{\psi}_{ab}, \qquad
\d_S \hat{\psi}_{cd} = 0.
\ee
where $\psi_{ab}$ is a new field that has only one holomorphic
component, $\psi_{\bar{z} \bar{z}}(z)$. (To extend this discussion
to the anti-holomorphic sector we would need to also turn on
$\tilde{\psi}_{zz}(\bar{z})$, i.e. the full transformation
is $\d_S g_{ab} = P^{(-)c}_{\ \ a} P^{(-)d}_{\ \ b} \psi_{cd} +
P^{(+)c}_{\ \ a} P^{(+)d}_{\ \ b} \tilde{\psi}_{cd}$).

Since the metric now transforms,
the action is not invariant and its $Q_S$ variation yields,
\be
\d_S S_\s = - \frac{1}{2} \int d^2 \s \sqrt{g}\ T^{ab} \d_S g_{ab}
=- \frac{1}{2} \int d^2 \s \sqrt{g} g^{ac} g^{bd} T^B_{ab} \hat{\psi}_{cd},
\ee
where again we only discuss the holomorphic sector, and
in the second equality we used the fact that due to
the projector operators the second term in (\ref{e/m1})
does not contribute. To construct an invariant action we now add
a new term to the action,
\be \label{t_ac}
S_\s \to S=S_\s + \frac{1}{2} \int d^2 \s \sqrt{g} g^{ac} g^{bd}
G_{ab}  \hat{\psi}_{cd}
\ee
The new action would be invariant provided there exists $G_{ab}$
transforming as
\be \label{G}
\d_S G_{ab} =  T^{B}_{ab}
\ee
Note that because $\hat{\psi}_{ab}$ has only one fermionic
component, the variation of the explicit worldsheet metrics
in the new term does not contribute. Including both sectors one finds
that for the discussion to go through $G_{ab}$ must be traceless.
Equation (\ref{G}) for $G_{ab}$
is precisely the equation for a composite ``b-field''. Such a
composite field has been constructed in conformal gauge
and one may covariantize it to obtain a $Q_S$, diffeomorphism and Weyl
invariant action. We will come back to the solution of (\ref{G}) later on.

\section{Adding vertex operators} \label{VO}

We will be interested in computing scattering  amplitudes. For this aim,
it is useful to introduce sources
$\r^i$ with Weyl weight one that couple to vertex operators $V_i$
that are scalar functionals with Weyl weight minus one \cite{Craps:2005wk}.
Then our starting point is the extended action
\be \label{S}
{\cal S} = S + \sum_{i=1}^n \r^i V_i[\varphi] (\s_i, \zeta_i)
\ee
where $S$ is given in (\ref{t_ac}), $\varphi$ denotes
collectively all worldsheet fields and we will shortly discuss
the vertex operator. The sources $\r^i$ are considered
to be infinitesimal, i.e. we only differentiate once with respect to
each source and then set them to zero. The new action (\ref{S})
depends on the positions of the vertex operators $\s_i^a$ and their
$Q_S$ partners $\zeta_i^a$,
\be
\d_S \s_i^a = \z_i^a, \qquad \d_S \z_i^a =0\, ,
\ee
or in complex coordinates,
\be
\d_S z_i = \z_i, \quad \d_S \bar{z}_i = \bar{\z}_i, \qquad
 \d_S \z_i = 0, \quad \d_S \bar{\z}_i = 0\, .
\ee
In keeping with the discussion of the previous sections we
will mostly focus on the holomorphic sector.
The positions  $\s_i^a$ and $\zeta_i^a$ are regarded as
new constant fields which we integrate over
in the path integral. This is somewhat unconventional but as demonstrated
in \cite{Craps:2005wk} for the case of the bosonic string
it allows for a uniform derivation of scattering
amplitudes with integrated and unintegrated vertex operators.
Here we extend that discussion to include the fermionic coordinates
$\z_i^a$.
The action (\ref{S}) is invariant under diffeomorphisms provided
one transforms the position $\s_i^a$ of the vertex operator $V_i$
and of $\z_i^a$ appropriately (the corresponding BRST transformations
are given  in  (\ref{pos_brs})).
Furthermore, we need to ensure that (\ref{S}) is $Q_S$ invariant.
Since $\z_i$ is a fermionic variable $V_i$ has
the expansion (in complex basis)
\be
V_i[\varphi](z_i, \z_i)
= V_i^{(0)}[\varphi](z_i) + \z_i V_i^{(1)}[\varphi](z_i)\, ,
\ee
where again we focus on the holomorphic sector.
For (\ref{S}) to be $Q_S$ invariant, we need
\be
\d_S \left(V_i[\varphi](z_i,\z_i)\right)=0\, .
\ee
The $Q_S$ transformation can act either on worldsheet fields $\varphi$
or on the positions $z_i$ and we obtain
\be \label{vi}
\d_S V_i[\varphi](z_i,\z_i) = (\d_S V_i^{(0)})(z_i) + \z_i
\left( \pa V_i^{(0)}(z_i) - (\d_S V_i^{(1)})(z_i) \right)
\ee
which implies
\be \label{vi2}
\d_S V_i^{(0)}=0, \qquad \d_S V_i^{(1)} = \pa V_i^{(0)}\, ,
\ee
where now $Q_S$ acts only on the fields.
>From (\ref{vi2}) we find that the integrated vertex operator
\be
U_i = \int dz V_i^{(1)}
\ee
is $Q_S$ invariant.

\section{BRST quantization} \label{BRST}

The action (\ref{S}) constructed in the previous section is invariant
under diffeomorphisms and local Weyl transformations. We will now
proceed to quantize this system using standard BRST methods.
As in \cite{Craps:2005wk, Becchi:1995ik}, our BRST analysis includes the
``gauge invariances'' due to zero modes. This is done using the
Batalin-Vilkovisky (BV) quantization scheme
\cite{Batalin:1984jr,Henneaux:1992ig}. Our treatment is a straightforward
extension of the analysis in \cite{Craps:2005wk}, so we will mostly
quote results; for a detailed discussion of the method including  a concise
self-contained summary of BV we refer to \cite{Craps:2005wk}.

We introduce diffeomorphism and Weyl ghosts, $c^a$ and $C_{\omega}$,
and their $Q_S$ partners,
\be
\d_S c^a = \g^a, \qquad \d_S C_{\w} = \g_{\w}\, .
\ee
The BRST transformations of all fields are given as usual by replacing the
gauge parameter by the corresponding ghost. We will need below the
explicit transformation of the metric and its $Q_S$ partner,
\be
\delta_V g_{ab}=\mathcal{L}_c g_{ab}+2C_{\omega}g_{ab},  \qquad
\delta_V
\hat{\psi}_{ab}=\mathcal{L}_c\hat{\psi}_{ab}-\mathcal{L}_{\gamma}g_{ab}
-2\gamma_{\omega}g_{ab}+2C_{\omega}\hat{\psi}_{ab}
\ee
and of the positions of the vertex operators and their
 $Q_S$ partners,
\be \label{pos_brs}
\delta_V \sigma_i^a=-c^{a}(\sigma_i), \qquad
\delta_V \zeta^{a}_i=\gamma^a(\sigma_i).
\ee
As discussed in \cite{Craps:2005wk}, the zero modes of the ghost fields
are associated with a gauge invariance of the ghost action
which should be gauge fixed.
Following the BV quantization scheme,  one should  introduce
ghost-for-ghosts, extraghosts, antighosts and associated
auxiliary fields to gauge fix this invariance.  For the problem at hand,
(as explained in  \cite{Craps:2005wk})
all these ``fields'' are constant, i.e. do not depend on the
worldsheet coordinates,  ghosts-for-ghosts are not needed and the
metric moduli $\t^k, k=1,...,6 g -6, \ (g \geq 2)$,
(considered as constant fields) play the role of extraghosts
\footnote{
Earlier works where the moduli
were treated as quantum mechanical degrees of freedom include
\cite{Baulieu:1987tq,Labastida:1988zb,Mansfield:1993qd,Becchi:1995ik}.}.
Recalling that due to the $Q_S$ symmetry all fields come in $Q_S$-multiplets,
we end up introducing the following constant fields
and BRST transformations,
\be
\delta_S \tau^k=\hat{\tau}^k, \qquad
\delta_V \tau^k=\xi^k, \qquad \delta_S\xi^k=\hat{\xi}^k,\qquad
\delta_V \hat{\tau}^k=-\hat{\xi}^k.
\ee
We further need antighost fields and corresponding
auxiliary fields
\be
\delta_S \tilde{\beta}^{ab}=\tilde{b}^{ab}, \qquad
\delta_V \tilde{\beta}^{ab}=-p^{ab}, \qquad
\delta_V \tilde{b}^{ab}=\pi^{ab}, \qquad
\delta_S p^{ab}=\pi^{ab}
\ee
(which in our conventions are tensor densities).

To gauge fix diffeomorphism and Weyl transformations we set the
worldsheet metric $g_{ab}$ equal to a reference metric
$\hat{g}_{ab}(\tau)$.
This can be implemented in the path integral via the following
gauge fixing Lagrangian,
\bea \label{L1}
 L_{1} &=&\delta_V \delta_S (\tilde{\beta}^{ab}[g_{ab}-\hat{g}_{ab}(\tau)])
\\
&=&
\delta_V(\tilde{b}^{ab}[g_{ab}-\hat{g}_{ab}(\tau)]
+\tilde{\beta}^{ab}[\hat{\psi}_{ab}-\hat{\tau}^k\partial_k\hat{g}_{ab}(\tau)])
\nonumber \\
&=&\pi^{ab}[g_{ab}-\hat{g}_{ab}(\tau)]
-\tilde{b}^{ab}[2C_{\omega}g_{ab}
+\mathcal{L}_cg_{ab}-\xi^k\partial_k\hat{g}_{ab}(\tau)]
-p^{ab}[\hat{\psi}_{ab}-\hat{\tau}^k\partial_k\hat{g}_{ab}(\tau)]
\nonumber \\
&+&
\tilde{\beta}^{ab}[\mathcal{L}_c \hat{\psi}_{ab}
+2 C_{\omega} \hat{\psi}_{ab}
-\mathcal{L}_{\gamma}g_{ab}-2\gamma_{\omega}g_{ab}
+\hat{\xi}^k\partial_k \hat{g}_{ab}(\tau)
-\hat{\tau}^k\xi^l\partial_k\partial_l\hat{g}_{ab}(\tau)]\, , \nonumber
\eea
where $\partial_k\hat{g}_{ab}(\tau)
= \partial \hat{g}_{ab}(\tau)/\partial \tau^k$ is the derivative
of the reference metric w.r.t. the moduli and $\hat{\psi}_{ab}$
is defined in (\ref{Sgab}).
This gauge fixing action contains the usual gauge fixing terms
for the metric and the ghost actions for $\tilde{b}$, $c$ and
$\tilde{\beta},\gamma$.

In addition when the Riemann surface has $\k$ conformal Killing
vectors\footnote{Recall that $\k=6$ for
a Riemann surface of genus 0,
$\k=2$ for genus one and $\k=0$ for higher genus surfaces.}
we need to fix $\k$ additional constant ``gauge'' symmetries. This is
done by fixing the position of $\k$ vertex operators; we call this
set $f$. From now on $\sigma^a_i$ will always belong to the complement
of $f$ and $\sigma^a_{\hat{i}}$ to $f$.
We further introduce additional constant antighosts and
associated auxiliary fields,
\be
\delta_S \beta^{\hat{j}}_a=b^{\hat{j}}_a,, \qquad
\delta_V \beta_a^{\hat{j}}=-p_a^{\hat{j}}, \qquad
\delta_V b^{\hat{j}}_a=\pi^{\hat{j}}_a, \qquad
\delta_S p_a^{\hat{j}}=\pi^{\hat{j}}_a \qquad (a,\hat{j}) \in f
\ee
and the following gauge fixing term
\bea \label{L2}
L_{2}&=&\delta_V \delta_S
\left(\sum_f \beta^{\hat{j}}_a(\sigma^a_{\hat{j}}
-\hat{\sigma}^a_{\hat{j}})\right)
=\delta_V \left(\sum_f b^{\hat{j}}_a(\sigma^a_{\hat{j}}
-\hat{\sigma}^a_{\hat{j}})+\beta^{\hat{j}}_a \zeta^a_{\hat{j}}\right)
\nonumber \\
&=&\sum_f \pi^{\hat{j}}_a(\sigma^a_{\hat{j}}-\hat{\sigma}^a_{\hat{j}})
-b^{\hat{j}}_a c^a(\sigma_{\hat{j}})-p^{\hat{j}}_a
\zeta^a_{\hat{j}}+\beta^{\hat{j}}_a\gamma^a(\sigma_{\hat{j}})\, .
\eea
At this point we have treated all gauge symmetries, except the
ones associated with zero modes of the original fields
$X, p, \theta, w, \l$. We will discuss these in the next section.

To summarize, the generating functional of scattering amplitudes
is given by
\be \label{gen}
Z[\s_i;\r^i] =
\int d \mu_\s d \mu \exp \left(- {\cal S} - L_1 - L_2\right)
\ee
where ${\cal S}, L_1$ and $L_2$ are given in (\ref{S}),(\ref{L1})
and (\ref{L2}), $d \mu_\s$ is the
measure factor associated with $X, p, \theta, w, \l$ (and
non-minimal variables) that we will discuss in the next section
and $d\mu$ is the measure that follows from the
analysis of this section, i.e.
\bea
d \mu &=&
\prod_i^n d^2 \s_i \sqrt{g(\s_i)} d^2 \z_i
\prod_{k=1}^{6 g-6} d \t^k d \xi^k d \hat{\t}^k d \hat{\xi}^k
\prod_{f}db^{\hat{j}}_a dp^{\hat{j}}_a d\beta^{\hat{j}}_a d\pi^{\hat{j}}_a
\times \nonumber \\
&\times& [d \psi_{ab}] [dg_{ab}][dc^a] [d\gamma^a]
[dC_{\omega}][d\gamma_{\omega}][dp^{ab}][d\tilde{\beta}^{ab}]
[d\pi^{ab}][d\tilde{b}^{ab}]
\eea
The first line contains the integration over all constant ``fields''
while the second line the fields we functionally integrate over.
The integration over most of these variables can be done exactly
as we now discuss.

As in previous sections we only discuss
the holomorphic sector. Firstly, integrating over $\pi^{ab}$
and $g_{ab}$ sets the worldsheet metric equal to the reference metric
$\hat{g}_{ab}$ in all expressions.
Integrating over $\pi_a^{\hat{j}}, p_a^{\hat{j}}$, leads to
delta functions
$\d (z_{\hat{j}}-\hat{z}_{\hat{j}}) \d (\z_{\hat{j}})$
which can be used to integrate over $z_{\hat{j}}, \z_{\hat{j}}$.
So $\k$ insertions\footnote{In the holomorphic section $\k=3$
for a Riemann surface of genus 0,
$\k=1$ for genus one and $\k=0$ for higher genus surfaces.}
 will involve
$V_{\hat{j}}^{(0)}(\hat{z}_{\hat{j}})$ while the remaining $(n-\k)$ vertex
operators will involve $V^{(1)}_i(z_i)$ and will be integrated.
Furthermore integrating out $b^{\hat{j}},  \beta^{\hat{j}}$
leads to the insertion $c(\hat{z}_{\hat{j}}) \d (\g(\hat{z}_{\hat{j}}))$.

Note that the $V_{\hat{j}}^{(0)}$ and $V_{i}^{(1)}$ do not
depend on the ghost fields, so the path integral factorizes
into a part that only depends on the ghosts and the
rest. One might anticipate that the ghost contributions
will cancel each other
since $c^a, C_{\w}$ and the $\g^a,\g_{\w}$ are related by
the $Q_S$ symmetry. So to simplify the presentation we
set to zero the ghosts. The complete computation
including the ghosts is given in appendix \ref{ghosts}.
The scattering amplitudes thus take the form
\be \label{sc}
\langle V_1 \cdots V_n \rangle =
\int d \mu_\s e^{-S_\s} d \tilde{\mu} e^{-\tilde{S}}
\prod_{\hat{j}=1}^\k V_{\hat{j}}^{(0)}(\hat{z}_{\hat{j}})
 \prod_{i=\k+1}^n \int d z_i V_{i}^{(1)} (z_i),
\ee
where
\be \label{ppsi}
d \tilde{\mu}  e^{-\tilde{S}}
= \prod_{k=1}^{6 g-6} d \t^k d \hat{\t}^k [d \psi_{ab}] [d p^{ab}]
\exp\int d^2 \s\left(
\sqrt{\hat{g}} \frac{1}{2} G^{ab}  \hat{\psi}_{ab} +
p^{ab} [\hat{\psi}_{ab}-\hat{\tau}^k\partial_k\hat{g}_{ab}(\tau)]\right)
\ee
Integrating out $p^{ab}$ gives a delta function that sets
$\hat{\psi}_{ab} = \hat{\tau}^k \pa_k \hat{g}_{ab}(\tau)$.
Finally integrating out $\hat{\t}^k$ leads to $(6 g - 6)$
(of which $(3 g-3)$ are holomorphic) insertions of $G^{ab}$,
\be \label{f_amp}
\langle V_1 \cdots V_n \rangle =
\int d \mu_\s  e^{-S_\s}
\prod_{k} d \t^k (G,\pa_k \hat{g})
\prod_{\hat{j}=1}^\k
V_{\hat{j}}^{(0)}(\hat{z}_{\hat{j}})
 \prod_{i=\k+1}^n \int d z_i V_{i}^{(1)} (z_i)
\ee
where $(G,\pa_k \hat{g}) = \int_{\Sigma} d^2 \s \sqrt{\hat{g}}
G^{ab} \pa_k \hat{g}_{ab}$.

\subsection{Summary}

Let us summarize the results so far. We started from a theory
with a fermionic nilpotent symmetry $Q_S$ and zero central charge
and we coupled it to topological gravity in a way that preserves
the $Q_S$ symmetry. Quantizing this system using standard BRST-BV methods
leads to the formula (\ref{f_amp}) for the scattering amplitudes.
In this formula the position of $\k$ of the vertex operators
$V_{i}^{(0)}$ is fixed while the remaining ones, $V^{(1)}_i$, are integrated.
These vertex operators satisfy (in the holomorphic sector),
\be
\d_S V_i^{(0)}=0, \qquad \d_S V_i^{(1)} = \pa V_i^{(0)}\, .
\ee
Furthermore, one needs $(6 g-6)$ insertions ($(3 g -3)$ holomorphic ones)
of the field $G_{ab}$ defined by
\be
\d_S G_{ab} = T_{ab}
\ee
where $T_{ab}$ is the stress energy tensor of the worldsheet theory.
This composite field is the  analogue of the $b$-antighost
in the scattering prescription of bosonic string theory.
One may have anticipated these results based on the scattering amplitude
prescription for the bosonic string and studies of topological strings.
Indeed this is precisely the prescription used in the literature.
The novelty here is its derivation from a first principles BRST-BV
quantization. Notice that these results hold irrespectively of
what the original sigma model is. In the next
section we discuss issues specific to the pure spinor
theory described in section \ref{review}.

\section{Pure spinor measure}\label{ps_measure}

We now return to the pure spinor sigma model. We would like to
understand the path integral measure $d \mu_\s$ and find the
explicit form of $G_{ab}$. The path integral measure will be
derived by gauge fixing ``invariances due to zero modes'',
as in the previous section. There is an important difference
however. The vertex operators in general depend on all fields
$X, \theta, \pi, w, \l$,
so the zero modes imply only an invariance of the action $S_\s$ in
(\ref{fl_action}) and not of the generating functional of
correlators  in (\ref{gen}). At first sight it seems as though one need
not gauge fix this invariance of $S_\s$. Indeed fermionic
zero modes do not present a problem; the vertex operators can provide
the appropriate number of fermionic zero modes so that the final
expressions are non-vanishing. Non-compact
bosonic zero modes however are still a problem, even in the
presence of vertex operators, because typically integration over
them leads to a divergent path integral;  the action $S_\s$
does not contain a convergence factor because of the zero mode
gauge invariance. This can be remedied by gauge fixing the
bosonic zero mode gauge invariances, as we discuss in this
section. As we shall see, because of the $Q_S$ invariance,
part of the invariance due to fermionic zero modes is also fixed.

On a genus $g$ surface, a worldsheet scalar $\F$ has one zero mode $\F_0$
and a worldsheet vector $P$ has $g$ zero modes,
$P_0(z) = \sum_{I=1}^g P^I \omega_I(z)$, where  $\omega_I(z)$ are
the $g$ holomorphic Abelian differentials of first kind satisfying
$\int_{A_I} dz \omega_J = \d_{IJ}$ and
the contour integral is around the $g$ non-trivial A-cycles of
a genus $g$ surface. Note that $\F_0$ and $P^I$ are constants.
In our case and in the minimal formulation
we have 10 zero modes $x_0^m$, 16 zero modes $\theta_0^\a$ and
11 zero modes $\l_0^\a$ from the worldsheet scalars and
$16 g$ zero modes $d_\a^I, I=1,\ldots g$,
and  $11 g$ zero modes $w_\a^I$ from the worldsheet vectors.
Of these $x_0^m, \l_0^\a$ and $w^I_\a$
are bosonic. The treatment of the zero modes of $x^m$
is standard and will not be discussed here.
Furthermore, following earlier work
we will trade $w_\a$, which transforms under the gauge transformation
(\ref{ginv}),
for the gauge invariant variables,
\be
N_{mn} = \frac{1}{2} w_\a (\g_{mn})^\a{}_\b \l^\b, \qquad J = w_\a \l^\a
\ee
where $N_{mn}$ is the (contribution of the pure spinors to the)
Lorentz current and $J$ is the ghost generator. As discussed
in \cite{Berkovits:2006vi}, the pure spinor condition
implies enough relations between $N_{mn}$ and $J$ so that one
can express the 11 independent components of $w_\a$ in terms of $J$
and 10 component of $N_{mn}$.
In what follows the $11g$ zero modes of $N_{mn}, J$
will be denoted by $N_{mn}^I, J^I$.

The BRST transformations corresponding to the zero mode gauge invariance
are given by
\be
\d_V \l_0^\a = c^\a, \qquad \d_V \theta_0^\a = \g^\a, \qquad
\d_V d^I_\a = \g_\a^I, \qquad \d_V w^I_\a = c_\a^I,
\ee
where $c^\a, c_\a^I$ are constant fermionic ghosts and
$\g^\a, \g_\a^I$ are constant bosonic ghosts.
 The transformations for $\l_0^\a, w^I_\a$
require some explanation, since $\l^\a$ satisfy a
quadratic constraint and $w_\a$ has a gauge invariance.
These zero modes are most easily described in $U(5)$ variables
since the system in terms of $\l^+, \l^{ab}, w_+, w_{ab}$
is unconstrained and has no gauge invariance (see appendix \ref{u5}).
The BRST transformation is then given by shifting these variables
by their zero modes. Reversing the steps in appendix \ref{u5}
one may express $c^\a$ in terms of the 11 zero modes of
$\l^+, \l^{ab}$ and $c_\a^I$ in terms of the $11g$ zero modes
of $w_+, w_{ab}$. The arbitrariness due to the gauge invariance
(\ref{ginv}) is then eliminated by passing to the
gauge invariant variables $N_{mn}^I, J^I$.

To maintain $Q_S$ invariance we further require
\be
\d_S \g^\a = c^\a, \qquad \d_S c_\a^I = \g_\a^I
\ee
To gauge fix the bosonic invariances we introduce  constant
fermionic and bosonic antighost fields, $b_\a, \tilde{b}_\a$
each containing 11 independent components,
$\tilde{b}^{mnI}$, $b^{mnI}$, each containing
$10g$ independent components and $\tilde{b}^I, b^I$,
each containing $g$ components and corresponding auxiliary fields.
The $Q_V$ and $Q_S$ transformations of these fields are given by
\bea \label{tr_au}
&&\d_S b_\a = \tilde{b}_\a, \qquad \d_S b^{mnI} = \tilde{b}^{mnI}, \qquad
\d_S b^I = \tilde{b}^I \\
&&\d_V b_\a = -\pi_\a, \qquad \d_V \tilde{b}_\a = \tilde{\pi}_\a \qquad
\d_V b^{mnI} = -\pi^{mnI}, \nonumber \\
&&\d_V \tilde{b}^{mnI} = \tilde{\pi}^{mnI},
\qquad \d_V b^I = -\pi^I, \qquad \d_V \tilde{b}^I =  \tilde{\pi}^I
\nonumber \\
&&\d_S \pi_\a = \tilde{\pi}_\a, \qquad
\d_S \pi^{mnI} = \tilde{\pi}^{mnI}, \qquad
\d_S \pi^I = \tilde{\pi}^I \nonumber
\eea

To gauge fix the zero mode gauge invariances we now introduce the
following gauge fixing Lagrangian
\bea
L_3 &=& \d_V \d_S \left(b_\a \theta_0^\a + \sum_{I=1}^{g}
(b^{mnI} N^I_{mn} + b^I J^I) \right) \\
&=& \d_V \left(-b_\a \l_0^\a + \tilde{b}_\a \theta_0^\a
+ \sum_{I=1}^{g} ( \frac{1}{2} b^{mnI} (d^I \g_{mn} \l_0)
+ \tilde{b}^{mnI}  N^I_{mn} + b^I (d^I \l_0)
+ \tilde{b}^I (w^I \l_0))
\right) \nonumber \\
&=& \pi_\a \l_0^\a + \tilde{\pi}_\a \theta_0^\a
+ \sum_{I=1}^g \left(-\pi^{mnI} \frac{1}{2} d^I \g_{mn} \l_0 +
\tilde{\pi}^{mnI} N_{mn}^I - \pi^I d_\a^I \l_0^\a
+ \tilde{\pi}^I J^I \right) \nonumber \\
&&+b_\a c^\a + \tilde{b}_\a \g^\a
+ \sum_{I=1}^g \left( \frac{1}{2} b^{mnI} ( \g^I \g_{mn} \l_0 - d^I \g_{mn} c)
- \frac{1}{2} \tilde{b}^{mnI} ( c^I \g_{mn} \l_0 - w^I \g_{mn} c)
\right. \nonumber \\
&&\left. +b^I (\g^I \l_0 - d^I c) -  \tilde{b}^I (c^I \l_0 - w^I c)\right)
\nonumber
\eea
Integrating over $b^\a$ and $\tilde{b}^\a$ leads to delta functions
for $c^\a$ and $\g^\a$, which can be used to integrate out
$c^\a, \g^\a$. Integrating over
$b^{mnI}, b^I, \tilde{b}^{mnI}, \tilde{b}^I$ yields $11 g$
delta functions $\d(\g^I \g_{mn} \l_0) \d(\g^I \l_0)
(c^I \g_{mn} \l_0) (c^I \l_0)$. The same argument that implies
that one can trade the $11 g$ zero modes of $w_\a$ for $N^I_{mn}$ and $J^I$
also implies that the delta functions set to zero $c^I, \g^I$
(with Jacobians canceling between the $\g^I$ and
$c^I$ terms). So the zero-mode measure now becomes
\bea \label{ps_mea}
[d \m_\s]_{z.m.} &=& [d^{16} \theta_0] [d^{11} \tilde{\pi}]
[d^{11} \l_0] [d^{11} \pi] \prod_{I=1}^g [d^{16} d^I][d^{11} \pi_I]
[d^{11} \tilde{\pi}_I] [d^{11} N_I] \times \\
&&\hspace{-2cm}\times
\exp \left( \pi_\a \l_0^\a + \tilde{\pi}_\a \theta_0^\a
+ \sum_{I=1}^g \left(-\pi^{mnI} \frac{1}{2} d^I \g_{mn} \l_0 +
\tilde{\pi}^{mnI} N_{mn}^I - \pi^I d_\a^I \l_0^\a
+ \tilde{\pi}^I J^I \right)\right), \nonumber
\eea
where $[d^{11} \l_0]$ and $\prod_{I} [d^{11} N_I]$ are the
Lorentz invariant integration measures derived in
\cite{Berkovits:2004px}, whose explicit form we will not need.
Our focus here is on the factors coming from integrating
over $\pi, \tilde{\pi}, \pi^I, \tilde{\pi}^I$.

\subsection{Minimal formulation}

Recall that $\pi_\a$ and $\tilde{\pi}_\a$ have 11 independent
components each. One way to parametrize them is to write
\be
\pi_\a = p_i C^i_\a, \qquad \tilde{\pi}_\a = \tilde{p}_i C^i_\a, \quad
i=1, \ldots, 11
\ee
where $p_i, \tilde{p}_i$ and the independent
components and $C_i^\a$ is a constant matrix of rank 11. Then
$[d^{11} \pi] [d^{11} \tilde{\pi}]= \prod_i dp_i d\tilde{p}_i$
and  integrating over
$p^i$ yields $\prod_i \d(C^i_\a \l_0^\a)$, while integrating over
$\tilde{p}^i$ yields  $\prod_i C^i_\a \theta_0^\a$.
Putting it differently, one may have started with antighosts
and auxiliary field $b^i, \tilde{b}^i, p^i, \tilde{p}^i$ and
gauge fixing
condition $C_\a^i \l_0^\a =0$, for the invariance due
to the $11$ zero modes of $\l^\a$
and gauge fixing
condition $C_\a^i \theta_0^\a =0$ for the invariance due to 11 of the 16 zero
modes of $\theta$. Note that the insertions
can be combined into 11 insertions of the ``picture-lowering'' operator
\be
Y_C= C_\a \theta_0^\a \d(C_\a \l_0^\a)
\ee
Similarly, we parametrize the $10g$ independent components of $\pi^{mnI}$ and
$\tilde{\pi}^{mnI}$ as
\be
\pi^{mnI} = p^{jI} B^{mn}_{jI}, \qquad \tilde{\pi}^{mnI} =
\tilde{p}^{jI} B^{mn}_{jI}, \qquad j=1,\ldots, 10
\ee
where $p^{jI}, \tilde{p}^{jI}$ are the $10g$ independent components
and $B^{mn}_{jI}$ are constants. Integrating over
$p_{jI}, \tilde{p}_{jI}$ and $\pi_{I}, \tilde{\pi}_{I}$
leads to the insertions
\be \label{pc_r}
\prod_{I=1}^g\left( (d_\a^I \l_0^\a) \delta(J^I) \prod_{j=1}^{10}
\frac{1}{2} B_{Ij}^{mn} (d^I \g_{mn} \l_0) \d(B_{Ij}^{mn} N^I_{mn})\right)
=\prod_{R=1}^g Z_J(z_R) \prod_{P=1}^{10g} Z_{B_P}(w_P)
\ee
where we reassembled the insertions in terms of the ``picture-raising''
operators
\be \label{Z}
Z_B = \frac{1}{2} B^{mn} d \g_{mn} \l \d(B^{mn} N_{mn}), \qquad
Z_J = (\l^\a d_\a) \d(J)
\ee
inserted at positions $z_R, w_P$. Here we use the fact that
the non-zero modes in the r.h.s. of (\ref{pc_r})
do not contribute in any correlator \cite{Berkovits:2004px}.
These insertions correspond to gauge fixing conditions
$B^{mn}_{iJ} N_{mn}^I = 0, J^I=0$, for the
gauge invariance due to the
$11g$  $w_\a$ zero modes and $B_{Ij}^{mn} (d^I \g_{mn} \l_0)=0,
d_\a^I \l_0^\a=0$ for the gauge invariance due to
$11g$ of the $16 g$ zero modes of $d_\a$.
Note that the constants $C_\a^i, B^{mn}_{jI}$ enter through
a gauge fixing term, so by standard arguments correlation
functions do not depend on them and their presence does not
imply breaking of Lorentz invariance.

What is left is to discuss $G_{ab}$. As we shall see,
we only need to recall well known facts from the literature.
By definition,
$G_{ab}$ should satisfy (now in complex coordinates
and dropping the indices)
\be \label{Geqn}
\d_S G = T, \qquad
T = \frac{1}{2} \Pi^m \Pi_m + d_\a \pa \theta^\a - w_\a \pa \l^\a
\ee
Since $\d_S$ is nilpotent, this equation defines a cohomology class
$[G]$, i.e. solutions $G$ up to $\d_S$ exact terms.
A solution of (\ref{Geqn}) is given by \cite{Berkovits:2001us}
\be \label{g0}
G_0 = \frac{C_\a G^\a}{C_\a \l^\a}, \qquad
G^\a = \frac{1}{2} \Pi^m (\g_m d)^\a
- \frac{1}{4} N_{mn}(\g^{mn} \pa \theta)^\a
-\frac{1}{4} J \pa \theta^\a
-\frac{1}{4} \pa^2 \theta^\a,
\ee
for a constant spinor $C_\a$.
This expression also appeared in \cite{Matone:2002ft} as a twisted
worldsheet supersymmetry current. This solution is however
not acceptable because had we allowed for operators with behavior
$(C_\a \l^\a)^{-1}$
the $Q_S$-cohomology would be trivial. Indeed, consider the
field $\xi$
\be \label{xi}
\xi= \frac{C_\a \theta^\a}{C_\a \l^\a}, \qquad \d_S \xi=1.
\ee
Then any closed operator $V$ is also exact since
\be \label{xi_ex}
\d_S V =0 \qquad \Rightarrow \qquad V = \d_S (\xi V).
\ee
A related issue is that the positions of the poles of
$G_0$ are also the positions of the zeros of the
path integral insertions thus making the expressions
ill-defined.

One might hope to arrive at a well-defined expression by
finding a different representative of the cohomology class $[G]$
such that the poles in the new $G$ would cancel
against zeros in other path integration insertions.
Indeed, such a representative $G_1$ exists and it is given by
 $G_1=b_B/Z_B$, where $Z_B$ is the picture raising operator in
(\ref{Z}) and $b_B$
is the ``picture-raised b ghosts'' constructed
in \cite{Berkovits:2004px} by solving the equation,
\be
\d_S b_B = Z_B T.
\ee
It was shown in \cite{Oda:2004bg} that $G_1$ is in the same cohomology class
as $G_0$. Using this solution we find that the poles of $G_1$
indeed cancel against zeros coming from the picture raising operators.

Combining all ingredients we find that the multi-loop amplitude
should include $3 g-3$ insertions of $b_B$, $10g - (3g -3)$ insertions
of $Z_B$, $g$ insertions of $Z_J$ and 11 insertions of $Y_C$.
This is precisely the prescription proposed in \cite{Berkovits:2006vi}.

\subsection{Non-minimal formulation}

Let us now return back to (\ref{ps_mea}) and recall that
$\pi_\a$ and $\tilde{\pi}_\a$ are $Q_S$ partners,
$\d_S \pi_\a = \tilde{\pi}_\a$, see (\ref{tr_au}),
and each has 11 independent components.
These are precisely the properties of the
non-minimal variables $\bar{\l}_\a$ and $r_\a$, see section 2,
so one may identify
\be
\pi_\a = \bar{\l}^0_\a, \qquad \tilde{\pi}_\a = r^0_\a
\ee
where $\bar{\l}^0_\a, r^0_\a$ are the zero modes
of $\bar{\l}_\a$ and $r_\a$. Actually
since the non-minimal variables
are cohomologically trivial their non-zero modes
do not contribute to any observable and one may only keep
their zero modes. Recall also that the non-minimal
sector has a gauge invariance similar to (\ref{ginv})
(whose explicit form is not needed here) and the following
combinations are gauge invariant \cite{Berkovits:2005bt}
\bea
&&\bar{N}_{mn}= \frac{1}{2} (\bar{w} \g_{mn} \bar{\l} - s \g_{mn} r),
\qquad \bar{J} = \bar{w}^\a \bar{\l}_\a - s^\a r_\a, \nonumber \\
&&S_{mn} = \frac{1}{2} s \g_{mn} \bar{\l}, \qquad S = s^\a \bar{\l}_\a
\eea
The canonical momenta
$\bar{w}^\a$ and $s^\a$ have $11g$ zero modes each
which, as in the discussion of the minimal variables,
can be traded for $10g$ zero modes of $\bar{N}_{mn}^I$ and $S_{mn}^I$
and $g$ zero modes of $\bar{J}^I$ and  $S^I$.
Using the $Q_S$ transformations in (\ref{qs_nm}) one finds
\be
\d_S S_{mn}^I = \bar{N}_{mn}^I, \qquad \d_S S^I = \bar{J}^I.
\ee
We thus find that the fields $\bar{N}_{mn}^I, S_{mn}^I, S^I, \bar{J}^I$
have the same number of components  and the same $Q_S$ transformations
as $\pi^{mnI}, \tilde{\pi}^{mnI}, \pi^I, \tilde{\pi}^I$,
and we can thus identify them,
\be
\pi^{mnI} = \bar{N}^{mnI}, \qquad \tilde{\pi}^{mnI} = S_{mn}^I,
\qquad \pi^I = S^I, \qquad \tilde{\pi}^I = \bar{J}^I\, .
\ee
With these identifications the exponential factor in
(\ref{ps_mea}) is precisely the regularization factor ${\cal N}$
in \cite{Berkovits:2005bt} (up to inconsequential numerical factors).

It remains to discuss $G_{ab}$. This field was constructed in
\cite{Berkovits:2005bt} (with an elegant interpretation of
the construction in terms of
{\v C}ech cohomology given in \cite{Berkovits:2006vi})
\begin{equation}
G_B=
\frac{\bar{\lambda}_{\alpha}G^{\alpha}}{(\bar{\lambda}\lambda)}
+\frac{\bar{\lambda}_{\alpha} r_{\beta}
H^{[\alpha \beta]}}{(\bar{\lambda}\lambda)^2}
-\frac{\bar{\lambda}_{\alpha}r_{\beta}r_{\gamma}
K^{[\alpha \beta \gamma]}}{(\bar{\lambda}\lambda)^3}
-\frac{\bar{\lambda}_{\alpha}r_{\beta} r_{\gamma} r_{\delta}
L^{[\alpha \beta \gamma \delta]}}{(\bar{\lambda}\lambda)^4}
\end{equation}
where $G^\a$ is given in (\ref{g0}) and $H^{\a\b}, K^{\a\b\g}, L^{\a\b\g\d}$
are explicitly known but we will not need their detailed form here.
Note also  that  this field is cohomologically equivalent to $G_0$
\cite{Oda:2007ak}.
Combining all ingredients
we thus arrive at the prescription proposed in \cite{Berkovits:2005bt}.

Noticed that $G_B$ field has poles as
$\bar{\l} \l \to 0$ so one might wonder whether this prescription
suffers from the same
problems as the one using $G_0$. Indeed, there is a
non-minimal version of the argument
around (\ref{xi})-(\ref{xi_ex}). The corresponding non-minimal $\xi$ field
is \cite{Berkovits:2005bt}
\be
\xi_{nm} = \frac{\bar{\l}_\a \theta^\a}{\bar{\l}_\b \l^\b + r_\b \theta^\b}
\ee
This diverges as $(\bar{\l} \l)^{-11}$ so one must ensure that
no operators which diverge with this rate are allowed. A related issue
is that the path integral with the insertions just discussed
will diverge if the insertions diverge as fast as $(\bar{\l} \l)^{-11}$.
As discussed in \cite{Berkovits:2005bt,Berkovits:2006vi}
this can only happen for genus $g>2$ (since the pure spinor
measure converges as  $(\bar{\l} \l)^{11}$ and $G_B$ diverges as
 $(\bar{\l} \l)^{-3}$).
One way to deal with this issue is look for a different representative
$G_{(B,\e)}$ of the $Q_S$ cohomology class of $[G]$
which is less singular than $G_B$ as $\bar{\l} \l \to 0$.
A construction of such $G_{(B,\e)}$ is presented in \cite{Berkovits:2006vi}.
Using this $G_{(B,\e)}$ field
one then arrives at a prescription that in principle
works to all orders.

This solves the problem in principle. The actual construction
of $G_{(B,\e)}$ however is very complicated.
Given that the issues with singularities are related to the
$\bar{\l} \l \to 0$ limit, a different approach would be to
modify the gauge fixing condition for the pure spinor zero modes
such that they are fixed to a non-zero value. It would be
interesting to investigate if  such gauge fixing
can be implemented and whether it would lead to a simpler scattering amplitude
prescription.

\section{Conclusions} \label{concl}

We presented in this paper a derivation of the scattering amplitude
prescription of the pure spinor superstring from first principles.
Our results confirmed the prescriptions advocated
in \cite{Berkovits:2004px} and \cite{Berkovits:2005bt,Berkovits:2006vi},
show that these prescriptions are equivalent
and also suggest avenues for searching for a simpler
prescription.

We now summarize our approach.
We considered the pure spinor model (i.e. the Green-Schwarz-Siegel action
plus the pure spinor variables) as a ``matter''
sigma model with target space
ten dimensional superspace (with embedding coordinates
$X, \theta$) times the pure spinor space (with embedding coordinates
$\l$).
To construct a string theory we coupled this model
to two dimensional (topological) gravity and then quantized the resulting
theory.
One should contrast this approach with previous works where the
aim was to find a model with local symmetry which upon gauge fixing
would lead to the pure spinor model with $Q_S$ emerging as
the BRST operator and the pure spinors $\l$ as the corresponding ghosts.
For us $Q_S$ and $\l$ are part of the model {\it ab initio}
and the justification for starting with this model is that the
$Q_S$ cohomology gives the superstring spectrum.
To maintain the $Q_S$ symmetry and consistently quantize the model after
coupling to $2d$ gravity, the $Q_S$ symmetry had
to be extended to act on the gravitational sector and we showed that
$Q_S$ invariance requires the existence of a (composite) field $G$ whose
$Q_S$ variation is equal to the  $2d$ stress energy tensor.

This model was then quantized
using standard BRST techniques, introducing diffeomorphism ghosts,
their $Q_S$ partners, associated auxiliary fields etc. It turns out that
all variables one introduces in this process can be explicitly
integrated out resulting in a prescription for the scattering
amplitudes involving (as usual) a number of unintegrated and
a number of integrated vertex operators and $(3g-3)$ (complex) insertions
of the zero modes of $G$. This result holds in general for any
system with a nilpotent symmetry coupled to topological gravity.

Our analysis included a BRST treatment of the gauge invariances
due to zero modes;
the presence of a zero mode implies an invariance of the
action under a shift of the field by the corresponding
zero mode. To gauge fix these invariances we introduced constant
ghosts, antighosts and corresponding auxiliary fields. In the
presence of vertex operators some of these invariances are
lifted. Nevertheless, one must still gauge fix all (non-compact)
bosonic invariances
because their presence implies that the worldsheet
action does not provide the appropriate
convergence factor for the integration over them.
We carried out this analysis for the bosonic zero modes
of the pure spinor sigma model.
This led (among other things) to the introduction of
constant auxiliary fields needed to implement the gauge fixing conditions
in the path integral. Depending
on the parametrization of these fields one is led either to the
minimal \cite{Berkovits:2004px} or the non-minimal \cite{Berkovits:2005bt}
prescription for scattering amplitudes. In the latter case
the auxiliary fields can be identified with the non-minimal variables
(more precisely, the zero modes of the non-minimal variables, but
since these variables are cohomologically trivial
their non-zero modes do not contribute to any observable).
To complete the construction one needs the explicit form of the
composite ``$b$-field'' $G$. The relevant results in the literature
nicely fit with our analysis and we thus arrived at the
precise form of the scattering amplitude prescriptions
in  \cite{Berkovits:2004px} and \cite{Berkovits:2005bt}.

The most complicated part of the scattering amplitude
prescription is the construction of a composite ``$b$-field''
with appropriate singular behavior. Although the existence
of a completely satisfactory $G$ field is guaranteed
by the results of \cite{Berkovits:2006vi}, the actual construction
is very complicated. A possible avenue towards a simpler prescription
would be to look for different gauge fixing conditions for the
zero modes, instead of looking for less singular representatives
of $[G]$ as has been done so far. We hope to report on this and
related issues in the future.

\section*{Acknowledgments}

We would like to thank Nathan Berkovits and Carlos Mafra for discussions.

\appendix

\section{$U(5)$ variables and the $Y$-formalism} \label{u5}

We discuss in this appendix the use of $U(5)$ variables
and the $Y$-formalism. We start by relaxing the pure spinor condition
on $\l^\a$ and introducing a Lagrange
multiplier $l_m$ to impose it in the path integral.
The $(w,\l)$ part of the  action (\ref{fl_action}) thus now reads
\be \label{ac_ps}
S_{(w,\l)}
= \int d^2 z \left(w_\a \bar{\partial} \l^\a + l_m (\l \g^m \l) \right).
\ee
where $\l^\a$ is now an unconstrained chiral spinor.
This action has a gauge invariance,
\be \label{gauge_tr}
\d w_\a =  \L_m (\g^m \l)_\a, \qquad
\d l_m = \frac{1}{2} \bar{\partial} \L_m + (\L \g_m \l)
\ee
where $\L^m$ and $\L^\a$ are gauge parameters. The $\L^\a$ gauge invariance
follows from the Fierz identity
\be \label{Fierz}
(\l \g^m \l) \g_m \l =0
\ee
that holds for any spinor $\l$. The same identity also implies that the
gauge algebra is reducible; the gauge transformations are invariant under
the transformation
\be
\d \L^\a = \left((\l \g^n \l) \g_n^{\a \b} - 2 \l^\a \l^\b \right)
\tilde{\L}_\b \; ,
\ee
with $\tilde{\L}_\b$ a new gauge parameter. This transformation
has a gauge invariance of its own, etc. The full set of
reducibility conditions is discussed in the appendix of
\cite{Berkovits:2001rb} and in \cite{Berkovits:2005hy}.
One may proceed to quantize this system in a manifestly
Lorentz invariant fashion by introducing ghosts-for-ghosts etc
but we shall not discuss this here. Instead we will
use a Lorentz breaking gauge fixing condition.

Let $\G^m$ be the $SO(10)$ gamma matrices and let us define
\be
\G^{+}_{a} = \frac{1}{2}(\G^{2 a} + i \G^{2 a-1}), \qquad
\G^{- a} = \frac{1}{2}(\G^{2 a} - i \G^{2 a-1}),
\quad a=1,2,3,4,5.
\ee
The spinor representation can be built by treating $\G^{a-}$
as annihilation and $\G_{a}^{+}$ as creation operators,
where $a=1,\ldots, 5$. Let us define
\be
v_{a_1...a_n}=\Gamma^{+}_{a_1}...\Gamma^{+}_{a_n} |0\rangle,
\quad
v_+=\Gamma^{+}_{1}\Gamma^{+}_{2}\Gamma^{+}_{3}
\Gamma^{+}_{4}\Gamma^{+}_{5} |0\rangle\,
\quad
v_-=|0\rangle.
\ee
where $n=1,\ldots, 4$. A chiral spinor has components $\l^+, \l^{ab},
\l_a = \e_{abcde} \l^{bcde}/24$, which transform as
$\underline{1}, \underline{10}$ and $\underline{5}^*$
under the $U(5)$ subgroup of $SO(10)$. In these variables
only 5 of the 10 expressions $\l \g^m \l$ are non-trivial; the
other 5 are automatically equal to zero if the first 5 hold,
\bea
\l \g_a^{+} \l &=& 2 \l_a \l^+ + \frac{1}{4} \e_{abcde} \l^{bc} \l^{de}\, ,
\label{a+} \\
\l \g^{a-} \l & = & -2 \l_b \l^{ab}\, . \label{a-}
\eea
Using (\ref{a+}) one finds that (\ref{a-}) is automatically
satisfied so without loss of generality
we can set to zero the Lagrange multipliers $l_{a-}$.
The action is now invariant under
\be
\d w_\a =  \L^a(\g^{+}_a \l)_\a, \qquad
\d l^{a+} = \frac{1}{2} \bar{\partial} \L^a, \qquad a=1, \ldots, 5
\ee
This gauge transformation has rank 5, so one can gauge fix it
by requiring
\be
w^a =0.
\ee
Following standard steps (and expressing the gamma matrices
in the $U(5)$ basis) we find that corresponding ghost action is
\be
\int d^2z \left(\bar{C}_b
(\gamma^{+}_{a})^{b}_{\ \beta} \lambda^{\beta} C^a +w^{a} \pi_a \right)
= \int d^2z \left(\bar{C}_a \lambda^{+} C^a +w^{a} \pi_a \right)
\ee
where $\bar{C}_b, C^a, \pi_a$ are the corresponding antighost, ghost and
auxiliary fields.
 Integrating them out sets $w^a=0$ and inserts in the
path integral measure the factor $(\l^+)^5$. Furthermore,
integrating out $l^{a+}$ leads to the delta function
$\d( 2 \l_a \l^+ + \frac{1}{4} \e_{abcde} \l^{bc} \l^{de})$
which can be used to integrate out $\l_a$ (so we are left with the
11 independent components $\l^+, \l^{ab}$) and also results in the
insertion $(\l^+)^{-5}$ in the path integral measure, which cancel
the factor  $(\l^+)^{5}$ from the ghosts.
The end result
is that the action (\ref{ac_ps}) becomes the free action
\be
\int d^2 z (w_+  \bar{\partial} \l^+ + w_{ab} \bar{\partial} \l^{ab}),
\ee
with all factors coming from eliminating the $\underline{5}^*$ and gauge fixing
the gauge invariance canceling out.

>From this local description one should now pass to the
global picture by gluing together the local pieces. The
general theory is presented in \cite{Witten:2005px,Nekrasov:2005wg}
and the pure spinor case has been discussed in detail in
\cite{Nekrasov:2005wg}. In general, there may be worldsheet
and target space diffeomorphism anomalies that render the theory inconsistent.
These were shown to cancel in the pure spinor case if one would
excise the $\l=0$ point from the space of pure spinors \cite{Nekrasov:2005wg}.
Furthermore,
requiring consistent gluing should also fix the path integral
measure. Since the theory is non-anomalous this measure should
be the Lorentz invariant measure determined in
\cite{Berkovits:2006vi}\footnote{To verify this one should first
determine the measure in terms of $\l^+, \l^{ab}, w^-, w_{ab}$
requiring invariance of the measure when we move from one patch to
another and then rewrite the resulting measure in a way that
is manifestly Lorentz invariant. For the $w^-, w^{ab}$ variables
this would involve changing variables to $N_{mn}, J$. As far as
we are aware this computation has not appeared in the literature,
see however \cite{Nekrasov:2005wg}.} .

Finally, let us briefly discuss the $Y$-formalism
of \cite{Oda:2005sd,Oda:2005wu,Oda:2007ak}. In this case one
introduces a  constant pure spinor $v_\a$
and the following projector,
\be
K_\a{}^\b = \frac{1}{2} (\g^m \l)_\a (Y \g_m)^\b
\ee
where $Y_\a = v_\a/(v_\a \l^\a)$. This projector has rank 5 (since
$\Tr K =5$) and can be used to solve the pure spinor condition,
\be \label{lk}
(\l \g^m \l) = 0 \quad \Leftrightarrow \quad \l^\a K_\a{}^\b =0.
\ee
Furthermore,
\be
(1 - K)_\a{}^\b (\g^m \l)_\b =0,
\ee
so $(\g^m \l)_\b$ also has rank 5. This means that the gauge invariance
can eliminate 5 of the components of $w_\a$, which can be done
using the gauge fixing condition,
\be \label{kw}
K_\a{}^\b w_\b =0.
\ee
Following our earlier discussion, one should now implement these
steps in the path integral. Up to issues related to possible
path integral insertions that could result from the details of
the integration over $l_m$ and the ghost, this
should result in the $Y$ formalism.

\section{Ghost contribution} \label{ghosts}

We discuss in this appendix the computation of the
contribution of the ghost fields
to scattering amplitudes. We will compute
\be
Z_{m} = \int
d \tilde{\mu}  e^{-\tilde{S}}  d \m_{gh} \exp( -S_{gh})
\ee
where $d \tilde{\mu}  e^{-\tilde{S}}$ is given in (\ref{ppsi}),
\be
d\mu_{gh}=[d\tilde{\beta}^{ab}][d\tilde{b}^{ab}][dc^a][d\gamma^a]
[dC_{\omega}][d\gamma_{\omega}][d\xi^k][d\hat{\xi}^k]
\prod_{\hat{j}=1}^\k c^a(\hat{\s}_{\hat{j}}) \d (\g^a(\hat{\s}_{\hat{j}})
\ee
and
\bea
S_{gh} &=& \int_{\Sigma} \left(
2 \gamma_{\omega}\tilde{\beta}^{ab} \hat{g}_{ab}(\tau)
-2 C_{\w} (\tilde{b}^{ab}\hat{g}_{ab}(\tau)
- \tilde{\beta}^{ab}\hat{\psi}_{ab}) \right. \\
&&\left. + \tilde{b}^{ab}[\hat{\nabla}_a c_b + \hat{\nabla}_b c_a]
+\tilde{\beta}^{ab}[\hat{\nabla}_a \gamma_b + \hat{\nabla}_b \gamma_a]
+ \tilde{b}^{ab} \xi^k \pa_k \hat{g}_{ab}(\tau) \right. \nonumber \\
&& \left.-\hat{\psi}_{ab}
[\partial_c(\tilde{\beta}^{ab}c^c)-2\tilde{\beta}^{c(b}\partial_c c^{a)}]
-\tilde{\beta}^{ab}[\hat{\xi}^k\partial_k\hat{g}_{ab}(\tau)
-\hat{\tau}^k\xi^l\partial_k\partial_l\hat{g}_{ab}(\tau)]\right) \nonumber
\eea
where $\hat{\nabla}_a$ is the covariant derivative associated with
$\hat{g}_{ab}$.

Integrating out $\gamma_{\omega}$ and $\beta(\tau) \equiv
\hat{g}_{ab}(\tau)\tilde{\beta}^{ab}$ sets the trace of
$\tilde{\beta}^{ab}$ equal to zero. We will denote by
$\beta^{ab}$ the traceless part of $\tilde{\beta}^{ab}$.
Integrating out $\hat{\xi}^k$  introduces
$(6 g -6)$ insertions of the $\b^{ab}$ zero modes,
while integrating
over $p^{ab}, \psi_{ab}$ and $\hat{\t}^k$ leads to insertions
of the zero mode of the ``supercurrent'',
\be \label{superc}
(\tilde{G}, \pa_k\hat{g}) \equiv \int_{\Sigma} d^2 \s
\left(\partial_c(\beta^{ab}c^c)
-2 \beta^{c(b}\partial_cc^{a)}
+2 \beta^{ab}C_{\omega}+\sqrt{\hat{g}}G^{ab}
+\beta^{ab} \xi^l \pa_l \right)
\partial_k \hat{g}_{ab}(\tau).
\ee

After these integrations we are left with
\be \label{zm}
Z_{m}
= \int d \mu_{\b\g} d \tilde{\mu}_{gh} \exp (- \tilde{S}_{gh})
\ee
where
\be
\tilde{S}_{gh} =
\int d^2 \s \left(\b^{ab} (\hat{\nabla}_a \g_b + \hat{\nabla}_b \g_a)
+ \tilde{b}^{ab}
(2C_{\omega}\hat{g}_{ab}+\hat{\nabla}_a c_b + \hat{\nabla}_b c_a)
+ \tilde{b}^{ab} \xi^k \pa_k \hat{g}_{ab}(\tau) \right)
\ee
and
\bea \label{mbg}
d \mu_{\b\g} &=&
[d \b^{ab}] [d \g^a]
\prod_{k=1}^{6 g -6} \d( (\b,\pa_k \hat{g})) \prod_{\hat{j}=1}^\k
\d(\g^a(\hat{\s}_{\hat{j}})) \nonumber \\
d \tilde{\mu}_{gh} &=&
[d\tilde{b}^{ab}][dc^a] [dC_{\omega}]
\prod_{k=1}^{6 g -6} d \t^k d \xi^k 
(\tilde{G},\partial_k\hat{g}(\tau))
\prod_{\hat{j}=1}^\k c^a(\hat{\s}_{\hat{j}})
\eea
The $\b\g$ system is now a standard CFT with a $U(1)$ ``ghost'' charge
conservation and the path integral measure contains all appropriate
zero mode insertions. It follows that the $\b$-dependent part of
(\ref{superc}) drops out of (\ref{zm}) since it is charged w.r.t. the $\b\g$
U(1). Integrating out $C_\w$ sets the trace of $\tilde{b}^{ab}$ to zero;
we will denote by $b^{ab}$ the traceless part,
and integrating out $\xi^k$ leads to $(6 g-6)$ insertions
of the $b^{ab}$ zero modes. We end up with
\be
Z_{m}
= \int d \mu_\t  d \mu_{\b\g} d \mu_{bc}
\exp\left(-\int d^2 \s \left(\b^{ab} (\hat{\nabla}_a \g_b
+ \hat{\nabla}_b \g_a) + b^{ab}
(\hat{\nabla}_a c_b + \hat{\nabla}_b c_a)\right)\right)
\ee
with $d \mu_{\b\g}$ as in (\ref{mbg}) and
\bea
d \mu_{bc}&=& [db^{ab}][dc^a] \prod_{k=1}^{6 g -6}
(b,\partial_k\hat{g}(\tau)) \prod_{\hat{j}=1}^\k c^a(\hat{\s}_{\hat{j}})
\nonumber \\
d \mu_\t &=& \prod_{k=1}^{6 g -6} d \t^k (G,\partial_k\hat{g}(\tau))
\eea
It is now manifest that the integration over $(b^{ab}, c^a)$
cancels against the integration over $(\b^{ab}, \g^a)$
and we are left with the same measure factor as in (\ref{f_amp}).

\end{document}